\documentclass[conference]{IEEEtran}

\usepackage[frozencache=true,cachedir=minted-cache]{minted}
\usepackage{booktabs}
\usepackage{multirow}
\usepackage{xspace}
\usepackage{hyperref}
\usepackage[capitalise]{cleveref}

\usepackage{subcaption}
\usepackage{todonotes}
\usepackage{csquotes}
\usepackage{url}
\usepackage{paralist}
\usepackage{footmisc}

\setminted{fontsize=\small}

\newcommand{\university}{University of Passau\xspace}
\newcommand{\toolname}{\emph{Perfumator}\xspace}

\newcommand{\summary}[2]{%
	\vspace{-0.2cm}%
	\begin{center}%
		\colorbox{gray!20}{%
			\parbox{\linewidth}{%
				\textbf{\textsf{Summary (\textit{#1})}:}~%
				#2%
			}%
		}%
	\end{center}%
}

\renewcommand{\paragraph}[1]{\noindent {\bf #1}}

\AtBeginDocument{%
	}

\begin{document}
	
	\title{Acknowledging Good Java Code\\ with Code Perfumes}

	\author{\IEEEauthorblockN{Philipp Straubinger}
			\IEEEauthorblockA{\textit{University of Passau} \\
					Passau, Germany}
			\and
			\IEEEauthorblockN{Florian Obermüller}
			\IEEEauthorblockA{\textit{University of Passau} \\
					Passau, Germany}
			\and
			\IEEEauthorblockN{Gordon Fraser}
			\IEEEauthorblockA{\textit{University of Passau} \\
					Passau, Germany}
		}
	
	\maketitle
	
	\begin{abstract}
		Java remains one of the most popular programming languages in
		education. Although Java programming education is well supported by
		study materials, learners also need more immediate support on the
		problems they face in their own code. When this support cannot be
		offered by educators personally, learners can resort to automated
		program analysis tools such as linters, which provide feedback on
		potential bugs or code issues. This is constructive feedback, but it
		may nevertheless feel like criticism.
		%
		This paper introduces \textit{code perfumes} for Java, a simple
		program analysis technique similar to linting, but commending the
		correct application of good programming practices. We present a
		catalogue of 20 Java code perfumes related to common Java language
		constructs for beginner to immediate learners.
		Our evaluation shows that these code perfumes occur frequently in
		learners' code, and programs with more code perfume instances tend
		to have better functionality and readability. Moreover, students who
		incorporate more code perfumes tend to achieve higher grades.
		Thus, code perfumes serve as a valuable tool to acknowledge
		learners' successes, and as a means to inform instructors about
		their learners' progress.
	\end{abstract}
	
	
	\section{Introduction}\label{sec:intro}

Java is one of the most frequently used languages to teach
programming~\cite{conceicao2022,santos2020InnovativeAI}, and teaching
materials and self-study materials are abundantly available. However,
independently of the maturity of these materials, students may
struggle in their attempts to produce functioning programs.  Even
when they produce code that implements the desired functionalities,
this is often prone to quality problems, which the students again may
struggle to identify and fix on their own~\cite{effenberger2022}.
In a self-study scenario, or when tutors are overwhelmed with large
classrooms and cannot provide help and feedback promptly, learners can
resort to automated program analysis tools such as
\textit{SpotBugs}\footnote{\url{https://spotbugs.github.io/}, last accessed
	08.06.24} or
\textit{SonarLint}\footnote{\label{sonar} 
\url{https://www.sonarsource.com/products/sonarlint/},
	last accessed 08.06.24}. Feedback provided by these tools on quality
issues can help learners produce better code and supports their
learning processes, and tutors benefit from insights into their
students’ current skills and problems or misunderstandings. However,
such automated feedback tools mostly focus on pointing out negative
aspects.

While this kind of corrective feedback was shown to be very helpful in
gaining further cognitive skills~\cite{wisniewski2020power}, purely
negative feedback may be detrimental to self-efficacy and decrease
intrinsic motivation~\cite{wisniewski2020power}. In contrast, positive
feedback is considered to have better effects on motivational
aspects~\cite{hattie2008visible}. Consequently, for effective learning
feedback should consist of information about errors as well as correct
behaviour to address both cognitive and motivational aspects.  To this
end the concept of \emph{code perfumes} was recently introduced for
the block-based Scratch programming
language~\cite{obermueller2021perfumes} as a means for automated
positive feedback: Code perfumes are patterns of code that are
considered as good programming practices or provide evidence of
correctly applied code constructs, implying an understanding of a certain
programming concept.

In this paper, we introduce the concept of code perfumes to the domain
of text-based programming languages by proposing a set of concrete
code perfumes for Java. For example,
the following code shows a student approach to
programming a scoring method in a chess game assignment.  The
first line in the method retrieves the colour of the human player from
the board. However, if the \mintinline{Java}{ChessBoard} object is
\mintinline{Java}{null} this method call results in a
\mintinline{Java}{NullPointerException}.
\begin{minted}{Java}
public static double score(ChessBoard board) {
  Color hColor = board.getHumanColor();
  // Method logic...
}
\end{minted}
A better solution is to use \emph{defensive programming} by checking
parameter objects for \mintinline{Java}{null} values before working
with them. The \emph{defensive null check} code perfume acknowledges
if learners correctly apply this approach by checking whether the
\mintinline{Java}{ChessBoard} object is \mintinline{Java}{null} at the
beginning of the method, and if so reacting with an
\mintinline{Java}{IllegalArgumentException}:
\begin{minted}{Java}
public static double score(ChessBoard board) {
  if (board == null) {
    throw new IllegalArgumentException(
    "Board is null!");
  }
  Color hColor = board.getHumanColor();
  // Method logic...
}
\end{minted}
We propose a total of 20 code perfumes covering various coding
concepts, and evaluate them on a dataset of 816 student submissions to
programming assignments in a university course. In detail, the
contributions of this paper are as follows:
\begin{itemize}
	\item We introduce the concept of code perfumes as a means of positive feedback to Java.
	\item We describe and implement a catalogue of 20 code perfumes from five categories.
	\item We empirically evaluate how frequently these code perfumes
	occur in student code.
	\item We empirically evaluate how the occurrence of code perfumes
	relates to correctness of the programs.
\end{itemize}

\noindent Our investigation shows that most of the proposed code
perfumes occur frequently in Java programming assignments.
Furthermore, our results suggest a relation between the occurrence of code
perfumes and correctness in the form of functionality, readability and
grades. These encouraging findings suggest that code perfumes can
provide positive feedback for Java and inform teachers about the
learning progress of students.


	\section{Background}\label{sec:background}

\subsection{Code Quality Problems}

Even though learning to program may be hard, most students will manage
to produce functioning code after some time.  However, research has
shown that even functionally correct programs show quality problems
and that students need feedback to stop producing
more~\cite{effenberger2022}.  For example, a common type of quality
problem is \emph{code smells}, which are code idioms that increase
the likelihood of errors in future program edits or decrease the
readability and understandability of the code~\cite{fowler1999}.
Prior research has shown that code smells have negative effects on
learning to program as they meddle with the learners' ability to
modify given code~\cite{hermans2016, michaeli2022}.  \emph{Bug
	patterns}, on the other hand, are aspects of code that are likely to
lead to undesired behaviour (i.e., bugs or
defects)~\cite{hovemeyer2004}.  In particular, misconceptions of
learners often manifest in bugs that follow recurring bug
patterns~\cite{swidan2018, fraedrich2020, sorva2018}.

While finding and removing such quality problems is inherent to coding
and learning to program, tools can offer support to learners as well
as teachers.  Static analysis tools (i.e., linters) like \textit{SpotBugs} and
\textit{SonarLint} can automatically detect these quality problems and generate
feedback, which has been shown to help students' code quality
skills~\cite{jansen2017impact}.  Furthermore, static analysis of code
patterns is part of automated feedback and grading tools like
ArTEMiS~\cite{DBLP:conf/sigcse/KruscheS18} providing information to
teachers for assessing their students.

\subsection{Feedback}

Feedback generated by automated program analysis tools is returned
regardless of other characteristics of the student and is accordingly
perceived as less threatening~\cite{hattie2008visible}.  No hand
raising is needed, thus preventing peer
stigmatisation~\cite{collins_kapur_2014}. Computer generated feedback
mainly focuses on bad coding practices or missing
functionalities~\cite{conceicao2022}.  Such negative feedback may be
useful in programming education in multiple ways: Teachers assessing
learners' code and supporting students with their individual problems
when programming need to know about their students' misunderstandings
and currently lacking skills and tools that can help in analysing
those~\cite{michaeli2019current,yadav2016expanding,sentance2017computing}.
Novices can benefit as corrective feedback helps to acquire further
knowledge and cognitive
skills~\cite{wisniewski2020power,jansen2017impact}.

On the other hand, positive feedback is important as it leads to
higher intrinsic motivation than negative
feedback~\cite{hattie2008visible}.  Previous research also suggests
that more motivated learners process all kinds of feedback
better~\cite{feedbackLearning}.  This in turn would possibly lead to
even higher effects of the corrective feedback. Unfortunately,
positive feedback generated by automated tools is
rare~\cite{conceicao2022}.  Most attempts to introduce positive
aspects to automated feedback are task-specific and give information
about what parts of a task are correct~\cite{martin2017}.  While this
is useful guidance on how to solve specified tasks, it gives no
information about general aspects, like style or quality.  One
attempt to address this deficiency in the context of Java is the use
of gamification for reducing warnings produced by analysis
tools~\cite{arai2014}. While this keeps motivation
high~\cite{zhan2022} and leads to better code, it only minimises the
negative feedback and does not contain any information about good
coding parts.

Positive feedback is more common in educational block-based
programming.  For the block-based programming language
Scratch~\cite{maloney2010} the Dr. Scratch~\cite{moreno2015} tool
displays praise in the form of a short text such as ``You're doing a great
job. Keep it up!!!'' and returns points on computational thinking
concepts\footnote{\url{http://www.drscratch.org/}, last accessed
	08.06.24}.  The LitterBox~\cite{fraser2021litterbox} static analysis
tool can not only point out bug patterns and code smells but also
detects \emph{code perfumes}, which are code idioms indicating good
programming practices or code showing understanding of certain
programming concepts~\cite{obermueller2021perfumes}. Code perfumes can
be seen as the opposite of code smells or bug patterns and are not
bound to a specific task but can be applied to any given project.
They have been shown to appear frequently in Scratch projects and can
be used as an indicator of correctness, as projects having a higher
count of code perfumes tend to be more
correct~\cite{obermueller2021perfumes}.

While Java is still one of the most used languages to learn
programming~\cite{conceicao2022,santos2020InnovativeAI}, similar
positive feedback is absent, which may limit the learning effects of
automated tools like ArTEMiS~\cite{DBLP:conf/sigcse/KruscheS18}. To
address this problem, we introduce code perfumes to this domain such
that holistic feedback combining positive and negative aspects can be
generated automatically.


	\section{Code Perfumes for Java}\label{sec:perfumes}

Code perfumes are a general concept applicable to any programming
language or skill level. In this paper, we target Java learners who
already have some basic understanding, and are able to produce code
using external APIs or design patterns. A specific concern related to
such learners is that even though they are familiar with the syntax,
they struggle to assess the quality of their
code~\cite{DBLP:conf/iticse/BorstlerSTADHJK17}. However, the code
perfumes we define may also appeal to more experienced programmers.

Code perfumes are expected to have mostly positive effects on the code
quality, be discernible in source code without execution, and align
with language-dependent concepts. However,
programmers must carefully assess their code and determine whether
applying a code perfume is appropriate in a specific context, or if
other solutions should be considered. This problem could be
exacerbated by false positives, which is why implementations of checks
for code perfumes need to aim to minimise false positives while
acknowledging the need for careful application.

We categorise code perfumes based on their purpose:
\begin{itemize}
	\item \textbf{Solution Patterns}: These are code structures that solve
	undesirable code patterns (e.g., code smells or bug patterns). They
	directly address and avoid matching negative patterns, but not all
	bug patterns imply solution
	patterns~\cite{DBLP:conf/wipsce/ObermullerBGH021}. Ideal solution
	patterns provide additional positive benefits beyond eliminating
	negative patterns.
	\item \textbf{Best Practices}: Established best practices in Java
	development can serve as the foundation for code perfumes. Capturing
	best practices in code patterns allow learners to leverage the
	results of experienced
	programmers~\cite{DBLP:conf/sigcse/PorterS13}.
	\item \textbf{Contracts or Conventions}: Java contracts, often
	documented in method descriptions, form another basis for code
	perfumes. Adhering to contracts, especially from the standard
	library is crucial for developing correct code and effective
	collaboration in development
	teams~\cite{DBLP:conf/aswec/DeveauxFJ01}.
\end{itemize}

To define code perfumes for Java, we draw inspiration from the
SonarRules collection of negative patterns by
SonarSource\cite{SonarLintRules} and
the popular book \enquote{Effective Java} by Joshua
Bloch~\cite{DBLP:books/lib/Bloch08}. SonarRules provides established
code smells, bug patterns, and vulnerabilities, influencing the
creation of \enquote{solution pattern} code
perfumes. \enquote{Effective Java} offers best practices and guidance
for Java programming, focusing on topics such as design patterns,
concurrency, and Java standard library traits. Additionally, the
Common Weakness
Enumeration\footnote{\url{https://cwe.mitre.org/index.html}, last accessed 
08.06.2024} was
considered as a background source, linking weaknesses identified by
SonarRules and providing additional information. Finally, the article
\enquote{Mining frequent bug-fix code
	changes}~\cite{DBLP:conf/csmr/OsmanLN14} also contributed to a code
perfume definition.

\begin{table*}[]
	\caption{Categories and types of code perfumes}
	\label{tab:perfumes}
	\resizebox{\linewidth}{!}{%
		\begin{tabular}{llll}
			\toprule
			\textbf{Category}                       & \textbf{ID} & \textbf{Code Perfume}                                  & \textbf{Type}                            \\ \midrule
			\multirow{7}{*}{Java Standard Library}  & 1           & Clone blueprint                                        & solution pattern, contract or convention  \\
			& 2           & Equals blueprint                                       & contract or convention, best practice    \\
			& 3           & Iterator next() follows the contract                   & contract or convention                   \\
			& 4           & Override compareTo with equals                         & solution pattern, best practice          \\
			& 5           & Override equals of superclass                          & solution pattern                         \\
			& 6           & Paired equals and hashCode                             & contract or convention, solution pattern  \\
			& 7           & Use optimized collections for Enums                    & best practice, solution pattern          \\ \midrule
			\multirow{6}{*}{Java language features} & 8           & ‘assert’ in private method                             & solution pattern, best practice          \\
			& 9           & At least X varargs                                     & solution pattern                         \\
			& 10          & Defensive default case                                 & solution pattern                         \\
			& 11          & Pattern matching with ‘instanceof’                     & solution pattern                         \\
			& 12          & Resource management in try-catch                       & best practice, solution pattern          \\
			& 13          & Synchronize accessors in pairs                         & solution pattern                         \\ \midrule
			\multirow{2}{*}{Unit Testing Practices} & 14          & JUnit 5 tests can be package-private                   & solution pattern                         \\
			& 15          & Single method call when testing for runtime exceptions & solution pattern                         \\ \midrule
			\multirow{2}{*}{Design Patterns}        & 16          & Builder pattern                                        & best practice                            \\
			& 17          & Singleton pattern                                      & best practice                            \\ \midrule
			\multirow{3}{*}{Others}                 & 18          & Copy constructor                                       & best practice                            \\
			& 19          & Defensive null check                                   & best practice, solution pattern          \\
			& 20          & No utility instantiation                               & solution pattern, best practice          \\ \bottomrule
		\end{tabular}
	}
\end{table*}

\subsection{Java Standard Library Code Perfumes}

Code perfumes in this category emphasise the importance of understanding
concepts from the Java standard library, which beginners should
encounter early in their coding journey~\cite{DBLP:books/lib/Bloch08}.

\paragraph{Clone blueprint:} This code perfume focuses on implementing the \mintinline{Java}{java.lang.Cloneable} interface to provide a copy functionality. While it may not be the ideal approach, it is a valid option if done correctly. The code perfume emphasises the importance of consistently calling \mintinline{Java}{super.clone} in \mintinline{Java}{clone} overrides in an inheritance hierarchy, to maintain the correct runtime typing of the cloned object. The code perfume aligns with the Common Weakness Enumeration (CWE-580) and the corresponding SonarSource rule (RSPEC-1182).
\begin{minted}{Java}
class DisplayData implements Cloneable {
  @Override
  public synchronized DisplayData clone() {
    DisplayData displayData;
    try {
      displayData = (DisplayData) super.clone();
    } catch (CloneNotSupportedException e) {
      throw new Error(e);
    }
    // Deep copy
    return displayData;
  }
}
\end{minted}

\paragraph{Equals blueprint:} Overriding the \mintinline{Java}{equals} method in the \mintinline{Java}{java.lang.} \mintinline{Java}{Object} class requires adherence to a specific contract, implementing an equivalence relation between instances of the class. To simplify the coding process and reduce the likelihood of errors, Joshua Bloch provides a guideline~\cite{DBLP:books/lib/Bloch08}. This code perfume focuses on the first two steps of Bloch's guideline, checking object references and type correctness, as the comparison of class fields can vary.
\begin{minted}{Java}
@Override
public boolean equals(Object o) {
  if (o == this) {
    return true;
  }
  if (!(o instanceof Pawn)) {
    return false;
  }
  // Field checks
}
\end{minted}

\paragraph{Iterator next() follows the contract:} This code perfume focuses on classes that implement the \mintinline{Java}{java.util.Iterator} interface in Java. This interface allows for iterating over data structures, such as lists. The key methods for iteration are \mintinline{Java}{hasNext} and \mintinline{Java}{next}, with the latter having a contractual requirement to throw a \mintinline{Java}{NoSuchElementException} when called without a remaining element in the iteration. This behaviour is expected by developers using iterators, and omitting it can lead to unexpected errors. This code perfume checks whether a class implementing the \mintinline{Java}{Iterator} interface adheres to the necessary step of throwing a \mintinline{Java}{NoSuchElementException} in the \mintinline{Java}{next} method. It is inspired by the rule from SonarSource (RSPEC-2272), which considers the absence of this behaviour a bug.
\begin{minted}{Java}
public final class Range 
  implements Iterator<Integer> {
  @Override
  public boolean hasNext() {
    return contains(start);
  }
  @Override
  public Integer next() {
    if (!hasNext()) {
      throw new NoSuchElementException();
    }
    // Iterator logic
  }
}
\end{minted}

\paragraph{Override compareTo with equals:} The \mintinline{Java}{java.lang.Comparable} interface includes the \mintinline{Java}{compareTo} method for defining a natural ordering between instances of the class. The method's contract suggests that if two objects are equal according to the \mintinline{Java}{equals} method, they should also be equal according to the \mintinline{Java}{compareTo} method, and vice versa. While this is only a recommendation, it is considered good practice to override both methods. The code perfume serves as a solution pattern to the corresponding SonarSource rule (RSPEC-1210) checking the presence one of these methods.
\begin{minted}{Java}
class Command implements Comparable<Command> {
  @Override
  public int compareTo(Command command) { ... }
  
  @Override
  public boolean equals(Object o) { ... }
}
\end{minted}

\paragraph{Override equals of superclass:} This code perfume addresses a common issue in inheritance hierarchies where subclasses add fields to those already defined in their superclasses. If a superclass overrides the \mintinline{Java}{equals} method, but the subclass fails to override it, there is a potential mistake. SonarSource (RSPEC-2160) identifies this, as it may lead to unexpected behaviour where the additional fields in the subclass are ignored by the \mintinline{Java}{equals} method.
\begin{minted}{Java}
public class Ancestor {
  @Override
  public boolean equals(Object o) { ... }
}
public class Child extends Ancestor {
  int additionalField;
  @Override
  public boolean equals(Object o) { ... }
}
\end{minted}

\paragraph{Paired equals and hashCode:} This code perfume highlights the importance of maintaining consistency between the \mintinline{Java}{equals} and \mintinline{Java}{hashCode} methods in Java. While classes can override these methods independently, it is considered best practice to override both or neither to avoid violating the contract they form. The contract, particularly for \mintinline{Java}{hashCode}, states that objects deemed equal by the \mintinline{Java}{equals} method should have the same hashCode. Although the contract, also denoted by SonarSource (RSPEC-1206) and CWE (CWE-581), allows distinct objects to have the same hashCode, it is recommended to override both methods together for consistent behaviour and improved performance in hash tables.
\begin{minted}{Java}
public class SudokuBoard implements Board {
  @Override
  public boolean equals(Object other) { ... }
  
  @Override
  public int hashCode() { ... }
}
\end{minted}

\paragraph{Use optimised collections for Enums:} This code perfume focuses on optimising memory usage and execution speed in Java applications, particularly when dealing with Enum types. It encourages the use of optimised classes provided by the Java standard library's \mintinline{Java}{Collections} API for Enums. The recommendation to use these optimised collections for Enums is emphasised by Joshua Bloch~\cite{DBLP:books/lib/Bloch08} and two SonarSource rules (RSPEC-1640 and RSPEC-1641).
\begin{minted}{Java}
public class KalahBoard implements Board {
  private final EnumMap<Player, PlayerPits> 
    playerPits;
  public KalahBoard( ... ) {
    this.playerPits = 
      new EnumMap<>(Player.class);
    // Constructor code
  }
}
\end{minted}

\subsection{Java Language Feature Code Perfumes}

Code perfumes in this category encourage the use of language features,
idioms, and keywords in Java coding practices, with a focus on
defensive programming~\cite{DBLP:conf/ACMse/TetoBL17} and addressing
code smells or bug patterns~\cite{DBLP:conf/sigcse/Weill-TessierCB21}.

\paragraph{‘assert’ in private method:}
While the \mintinline{Java}{assert} keyword is useful for validating
parameters, it should be avoided in public methods because assertions
must be explicitly enabled at the JVM level. Best practice involves
using \mintinline{Java}{assert} in private methods during development
for parameter validation. To enhance clarity and to ease error
identification, the \mintinline{Java}{assert} expression should be
accompanied by a message. This aligns with SonarSource's rule
(RSPEC-4274) against using \mintinline{Java}{asserts} to check public
method parameters, and is also highlighted in \enquote{Effective
  Java}~\cite{DBLP:books/lib/Bloch08}.
\begin{minted}{Java}
private boolean isMovePossible(int row, 
  int diag, Color color) {
  assert isValidPosition(row, diag) 
    : "Invalid coordinates";
  // Method logic
}
\end{minted}

\paragraph{At least X varargs:} This code perfume addresses the use of varargs in Java methods, emphasising a more efficient approach to set a lower bound on the number of parameters. Instead of checking the number of parameters within the method's body resulting in unnecessary code, it encourages declaring a specific number of parameters before the varargs. This approach, recommended by Joshua Bloch~\cite{DBLP:books/lib/Bloch08}, allows for compile-time validation of the minimum argument count, improving efficiency compared to runtime checks.
\begin{minted}{Java}
public static boolean isWithinCharset(
  int charsetSize, int... symbols) {
  return symbols != null 
    && Arrays.stream(symbols)
      .allMatch(i -> i >= 0 && i < charsetSize);
}
\end{minted}

\paragraph{Defensive default case:} This code perfume underscores the
importance of using \mintinline{Java}{switch} statements effectively
as a cleaner alternative to chaining
\mintinline{Java}{if}-statements. The \mintinline{Java}{switch}
construct allows handling every possibility of the
\mintinline{Java}{switch}-variable, with the
\mintinline{Java}{default} case serving as a safety net to address
unforeseen situations. Including a \mintinline{Java}{default}-case at
the end of a \mintinline{Java}{switch} statement demonstrates
defensive programming, reducing the risk of unexpected behaviour due
to overlooked cases as reported by SonarSource (RSPEC-131 and CWE
(CWE-478)).
\begin{minted}{Java}
switch (firstLetter) {
  case 'h':
    cmdHelp(token);
    break;
  // More cases
  default:
    printError("Invalid Command!");
    break;
}
\end{minted}

\paragraph{Pattern matching with ‘instanceof’:} This code perfume
promotes the use of the \mintinline{Java}{instanceof} operator in
conjunction with pattern matching, introduced in Java 16. While the
\mintinline{Java}{instanceof} operator traditionally required manual
casting, pattern matching now allows for a concise type-check and cast
in a single \mintinline{Java}{instanceof} statement. This enhances
clarity and readability, as promoted by SonarSource (RSPEC-6201).
\begin{minted}{Java}
if (copy instanceof Game gameCopy) {
  gameCopy.cells = cells.clone();
}
\end{minted}

\paragraph{Resource management in try-catch:} This code perfume encourages the use of the try-with-resources statement in Java to ensure the proper closing of resources within a try-catch block. The try-with-resources statement is considered a best practice inspired by SonarSource (RSPEC-2093) to prevent accidental resource leaks, especially for beginners working with file reading and I/O streams.
\begin{minted}{Java}
try (BufferedReader reader = new BufferedReader(
  new FileReader(file))) {
  Line line = readLine(reader, 0);
  // Method logic
}
\end{minted}

\paragraph{Synchronize accessors in pairs:} This code perfume encourages careful handling of parallelisation, emphasising the use of the \mintinline{Java}{synchronized} keyword to synchronise specific parts of the execution flow and prevent race conditions. It particularly focuses on scenarios involving accessor methods (getters and setters) for class fields. The code perfume rewards the practice of synchronising both accessors, addressing a potential bug as identified by SonarSource (RSPEC-2886).
\begin{minted}{Java}
synchronized void setCell(int rowNr, int colNr, 
  int number) { ... }

synchronized int getCell(int rowNr, int colNr) 
  { ... }
\end{minted}

\subsection{Unit Testing Code Perfumes}

Code perfumes in this category recognise the standard practice of unit
testing~\cite{DBLP:conf/issre/DakaF14} and the need for high-quality
test code to avoid maintainability
issues~\cite{ramler2016automated,martin2011clean,van2001refactoring}.

\paragraph{JUnit 5 tests can be package-private:} This code perfume advocates for adhering to the principle of operating with the least required privileges in Java coding. Specifically, it encourages developers to choose the narrowest scope possible for class visibility, including unit tests. With JUnit 5, both tests and test classes do not have to be public anymore, making them package-private is sufficient for JUnit to detect and execute them; inspired by SonarSource (RSPEC-5786).
\begin{minted}{Java}
class PerfumeTest {
  @Test
  void someTest() { ... }
}
\end{minted}

\paragraph{Single method call when testing for runtime exceptions:} This code perfume emphasises the importance of testing code for expected RuntimeExceptions using frameworks like JUnit and AssertJ. It discourages chaining multiple method calls within the \mintinline{Java}{assertThrows} or \mintinline{Java}{assertThatThrownBy} methods, as this can obscure which specific method call caused the expected exception or whether any call did at all. The code perfume suggests using only a single method call in these methods for improved code clarity and consistency as taken from SonarSource (RSPEC-5778).
\begin{minted}{Java}
@Test
void someTest() {
  assertThrows(Exception.class, 
    () -> callMethod());
}
\end{minted}

\subsection{Design Pattern Code Perfumes}

Code perfumes in this category acknowledge the value of design
patterns in advanced programming education and their contribution to
well-formed
code~\cite{helm2000design,fojtik2014design,DBLP:conf/kbse/BlewittBS05}.

\paragraph{Builder pattern:} This code perfume underscores the benefits of using the Builder pattern to address the challenge of classes with numerous optional fields. Instead of creating an excessive number of constructor overloads, the Builder pattern involves an inner class responsible for constructing instances of the outer class. This approach is more flexible, allowing easier extension of the builder and the creation of multiple instances using the same builder. Joshua Bloch~\cite{DBLP:books/lib/Bloch08} recommends considering the Builder pattern when a class's constructor would require four or more parameters.
\begin{minted}{Java}
public class ToBuild {
  private ToBuild(Builder builder) { ... }
  public static class Builder {
    public ToBuild build() {
      return new ToBuild(this);
    }
  }
}
\end{minted}

\paragraph{Singleton pattern:} This code perfume focuses on the Singleton pattern for enforcing the existence of exactly one instance of a class. Joshua Bloch~\cite{DBLP:books/lib/Bloch08} describes three main ways to implement the Singleton pattern: using a public instance field, a public static instance factory method, or defining the class as an enum type with a single constant. Each approach has its advantages, and all are considered valid.
\begin{minted}{Java}
static final class Controller {
  private static Controller instance;
  private Controller() { }
  public static Controller getInstance() {
    return instance;
  }
}
\end{minted}

\subsection{Other Code Perfumes}

Finally, some code perfumes fit in no other category.

\paragraph{Copy constructor:} This code perfume considers object copying in Java, emphasising the pitfalls associated with the \mintinline{Java}{Cloneable} interface, as highlighted by Joshua Bloch~\cite{DBLP:books/lib/Bloch08}. A copy constructor is recommended as a clean solution that takes an object of the same class as its parameter and performs a deep copy of all fields, including mutable non-primitive ones. This approach ensures the desired runtime type and is less error-prone than implementing \mintinline{Java}{Cloneable}.
\begin{minted}{Java}
public Tape(Tape clone) {
  content = new LinkedList<>(clone.content);
  headPosition = clone.headPosition;
  move(Direction.NOTHING);
}
\end{minted}

\paragraph{Defensive null check:} This code perfume underscores the importance of caution when declaring public methods that deal with non-primitive parameter types. The risk of bugs and unexpected behaviour is increased when null values are passed as parameters. To prevent unexpected \mintinline{Java}{NullPointerExceptions} at runtime, the code perfume recommends checking non-primitive parameters for non-null values or using annotations like \mintinline{Java}{@NotNull} or \mintinline{Java}{@Nonnull} to specify that null values are not allowed. Alternatively, if the method permits null values for certain parameters, using an \mintinline{Java}{@Nullable} annotation is advised by Joshua Bloch~\cite{DBLP:books/lib/Bloch08}.
\begin{minted}{Java}
public static void callPrint(
  TuringMachine turingMachine) {
  if (turingMachine != null) {
    System.out.println(turingMachine.print());
  } else {
    errorMsg("No Turing machine initialized.");
  }
}
\end{minted}

\paragraph{No utility instantiation:} This code perfume emphasises the definition of a private constructor for utility classes containing only static methods. Since utility classes do not need to be instantiated and thus need no implicit default constructor, having a single private constructor demonstrates an understanding of Java mechanics and object orientation as described by SonarSource (RSPEC-1118).
\begin{minted}{Java}
public final class Shell {
  private Shell() { }
  private static void process(String[] strings) 
    { ... }
  private static void add(String[] strings) 
    { ... }
}
\end{minted}

\subsection{\toolname: A Tool to Detect Code Perfumes in Java Code}

The \toolname is a static analysis tool implemented in Java 17, with its primary purpose being the detection of code perfumes in abstract syntax trees (ASTs). Its functionality involves two main steps: defining static information in a \mintinline{Java}{JSON} file and implementing a \mintinline{Java}{Detector} class that recognises specific patterns in the AST. JavaParser is utilized by the \toolname for parsing ASTs and resolving symbols. The tool offers a command-line interface featuring options like input path, output format, batch size, specifying dependencies as well as support for internationalization.

The \toolname identifies instances of code perfumes in designated input files or directories and generates serialised output in \mintinline{Java}{JSON} format containing detailed information. To enhance accuracy, the tool allows ignoring specific files and directories like the build directory during analysis. The \toolname is designed to be easily extendable with new code perfumes and holds the potential for integration into other tools.

	\section{Evaluation}\label{sec:evaluation}

To gain an initial understanding of the prevalence of the code
perfumes defined in this paper, and how they relate to correctness, we
applied the \toolname to a large dataset of programming assignments
and student solutions. This allows us to address the following
research questions:

\begin{itemize}
	\item \textbf{RQ1}: How common are code perfumes in Java?
	\item \textbf{RQ2}: Are code perfumes related to correctness?
\end{itemize}

\subsection{Experimental Setup}

\begin{table}[]
	\caption{Assignment submissions and size}
	\label{tab:submissions}
	\centering
	\begin{tabular}{lrrrr}
		\toprule
		Task & Name & Total LoC  & Submissions & Average LoC           \\ \midrule
		0    & Practice & 133931 & 359         & 373 			\\
		1    & Algorithms & 442450 & 809         & 547 			\\
		2    & CLI Board Game & 659318 & 607         & 1086 			\\
		3    & GUI Board Game & 609915 & 560         & 1089			\\ \bottomrule
	\end{tabular}
\end{table}

In order to study code perfumes, we use a dataset derived from the
\enquote{Programming II} course taught every semester at the
\university.  This course focuses on intermediate/advanced Java
programming, covering topics such as Java Swing and parallelism. The
coursework consists of four programming assignments.  Although the
specific tasks vary between different instances of the course, they
are always variations of the same type of program, and usually
represent the implementation of a board game like
Abalone\footnote{\url{https://en.wikipedia.org/wiki/Abalone_(board_game)}, last 
accessed 08.06.2024}
or
Mastermind\footnote{\url{https://en.wikipedia.org/wiki/Mastermind_(board_game)},
 last accessed 08.06.2024}. Due
to the similarity of the solution code, we merge different variants of
the same task in our dataset.
The tasks include:
\begin{inparaitem}
	\item Task 0: Exercise task for familiarity with the automatic grading system
	\item Task 1: Mathematical problem-solving to acquaint students with organising programs and creating complex algorithms
	\item Task 2: Modelling the board game with a command-line interface
	\item Task 3: Implementing a user interface for the previously modeled game
\end{inparaitem}

The first assignment only serves for practice, while the remaining
assignments determine the overall grade.  Until the submission
deadline, students can re-submit as often as they wish, receiving
feedback from a set of public tests. After the deadline, functionality
is assessed using a set of secret tests and normalised to the range 1
(all tests pass) to 6 (no tests pass). Additionally, each submission
is graded manually on readability considering coding styles, code
structure, and variable/method naming on a range of 1 to 6. The
overall grade is a weighted sum of functionality and readability
ranging from 1 (best) to 5 (worst) with intermediate steps of .3 and
.7.

We use data from this course between the winter semester of 2015/2016 to the
winter semester of 2022/2023, covering 15 iterations of the
course. Students who registered but did not submit a task were
excluded from the analysis. This resulted in a dataset of 816
students, acknowledging that some students may have participated more
than once to pass the course. \Cref{tab:submissions} shows that less
than half of the students submitted Task 0, which was optional. The
highest number of submissions was for Task 1, the first graded task,
with subsequent tasks seeing declining submission rates. The lines of
code (LoC) increased with each task, with Tasks 2 and 3 having almost
identical LoC counts.

To understand the limitations and applicability of our code perfumes,
we also consider the Progpedia dataset~\cite{paiva2023} consisting of
16 programming assignments for undergraduate students. The dataset
consists of 4362 student solutions. The assignments are very basic and
cover topics from reading user input from the console to basic
algorithms like Kruskal~\cite{kruskal1956shortest}.

\subsubsection{RQ1: How common are code perfumes in Java?}

To answer this research question, we execute the \toolname on all
student submissions for both datasets and extract the number of code
perfumes per category.

\subsubsection{RQ2: Are code perfumes related to correctness?}

To answer this research question, we compare the number of code
perfumes with other metrics of their solutions, such as the grade on
functionality, the grade on readability, the overall grades for the
course, and the number of lines of code (LoC), assessed using
\textit{CLOC}\footnote{\url{https://github.com/AlDanial/cloc}, last accessed 
08.06.2024}.
To contrast the insights gained from positive code aspects such as
code perfumes, we also examine negative aspects by extracting the
number of code smells using
\textit{SonarLint}\footref{sonar}. Subsequently,
we use Pearson correlation~\cite{cohen2009pearson} to determine
potential correlations between these metrics.

\subsection{Threats to Validity}

\emph{Threats to internal validity} may result from task selection,
grading criteria, and inconsistencies in both automatic and manual
grading processes. However, the grading process remained consistent
over the years with the same supervisor and the same grading
criteria. Moreover, the act of testing itself might influence
students' subsequent performance, and external factors like changes in
course structure or students' natural skill progression could also
impact the outcomes. The \toolname might not accurately detect all 
perfumes, but we minimized this issue by verifying the results beforehand.

\emph{Threats to external validity} exist because findings might not
generalise beyond the specific student population or programming
course and university. Additionally, the tasks may not accurately
mirror real-world programming scenarios, and changes over time in
technology, education methods, or student demographics could affect
the results.

\emph{Threats to construct validity} may result from the tools
utilised for assessment, which may not fully capture the intended
variables or could introduce errors. All students had the possibility
to analyse their graded results and discuss them with the supervisor,
therefore reducing this threat. Maintaining consistency in task
specifics and evaluation across different course iterations
potentially affects treatment effects.

\subsection{RQ1: How Common are Code Perfumes in Java?}

\begin{table}[t]
	\caption{Code Perfumes overall and per task}
	\label{tab:perfumestask}
	\centering
		\begin{tabular}{rrrrrrr}
			\toprule
			ID                                                                                        & Total & Submissions & Task 0 & Task 1 & Task 2 & Task 3 \\
			\midrule
			1                                                                                       & 196   & 4.0\% & 0      & 3      & 97     & 96     \\
			2                                                                                      & 17    & 0.3\% & 0      & 7      & 5      & 5      \\
			3                                                                  & 2     & 0.0\% & 0      & 0      & 1      & 1      \\
			4     																& 281   & 5.8\% & 0      & 254    & 12     & 15     \\
			5                                                                         & 0     & -- & 0      & 0      & 0      & 0      \\
			6         																& 419   & 8.3\% & 0      & 259    & 79     & 81     \\
			7                                                                   & 10    & 0.1\% & 0      & 0      & 5      & 5      \\
			8                                            								& 520   & 1.7\% & 1      & 6      & 257    & 256    \\
			9                                                                                    & 4     & 0.0\% & 0      & 1      & 3      & 0      \\
			10                                                                               & 3656  & 43.3\% & 426    & 871    & 1346   & 1013   \\
			11                                   								& 21    & 0.3\% & 0      & 2      & 2      & 17     \\
			12                                                                     & 134   & 3.0\% & 2      & 80     & 37     & 15     \\
			13                                                                       & 9     & 0.1\% & 0      & 0      & 0      & 9      \\
			14                                                                 & 0     & -- & 0      & 0      & 0      & 0      \\
			15                                               & 0     & -- & 0      & 0      & 0      & 0      \\
			16                                                                                      & 0     & -- & 0      & 0      & 0      & 0      \\
			17                                                                                    & 17    & 0.2\% & 0      & 0      & 6      & 11     \\
			18                                                                                     & 21    & 0.5\% & 0      & 5      & 7      & 9      \\
			19                                                                                 & 1048  & 11.5\% & 125    & 271    & 250    & 402    \\
			20                                                                             & 2243  & 43.0\% & 4303    & 851    & 671    & 418    \\ \bottomrule
		\end{tabular}
\end{table}

\Cref{tab:perfumestask} lists the overall number of code perfumes
found, their distribution across tasks, and the percentage of
submissions containing them.
In total, the \toolname identified 8598 instances of perfumes in the
dataset.
Only a few of the code perfumes are not found, which is due to the
nature of the tasks: The absence of the \enquote{override equals of
	superclass} code perfume (ID 5) is expected, given the relatively
simple project structures that may not involve extensive inheritance
hierarchies.
Code perfumes related to testing (IDs 14 and 15) are unsurprisingly
absent, considering the absence of unit testing in the
course. Similarly, design patterns (IDs 16 and 17) are infrequently
used, possibly due to students' limited awareness of builder and
singleton patterns or the tasks not necessitating their
implementation. The copy constructor code perfume (ID 18) sees limited
use, as most students opt for the cloning functionality (ID 1).
However, all other code perfumes can be found frequently in the
dataset, suggesting that they are relevant and suitable for the type
of code we are targeting.

The distribution of code perfumes reveals insights into the coding
behaviour of the students. For example, defensive programming
practices are very prevalent among students and these
assignments. Three specific code perfumes (IDs 10, 19, and 20)
collectively constitute over 80\% of all detected instances,
indicating a strong emphasis on defensive programming. Notably, the
\enquote{assert in private method} code perfume (ID 8) is frequently
encountered, along with the \enquote{defensive null check} code
perfume (ID 19), suggesting a preference for null checks in public
methods over assertions.

The code perfume occurrences also suggest adherence to certain Java
language conventions, particularly in cases where students override
the \mintinline{Java}{equals} method alongside either
\mintinline{Java}{hashCode} or \mintinline{Java}{compareTo} (IDs 4 and
6). However, the students notably deviate from the common
\mintinline{Java}{equals} blueprint by not consistently checking
object references and type correctness.
An intriguing observation is also that students rarely employ
synchronized accessors in pairs (ID 13), a crucial aspect introduced
in the final assignment of each course. This oversight could lead to
runtime exceptions and deadlocks in their implemented games.
Optimised collections for enums (ID 7) are infrequently used,
indicating potential unawareness of these specialised collections. The
same may apply to the \enquote{at least X varargs} code perfume (ID
9), while the \enquote{pattern matching with instanceof} code perfume
(ID 11) reflects the limited adoption due to its recent functionality
release in 2021.
This raises the question whether the inclusion of code perfumes into
student feedback could help improving such deficiencies.

It is noteworthy that the frequency of finding a code perfume does not
necessarily imply even distribution across all submissions. For
instance, although code perfume ID 20 is encountered twice as often as
code perfume ID 19, the former appears in nearly four times as many
submissions as the latter (see \cref{tab:perfumestask}). Similarly,
code perfumes 6 and 8 exhibit a similar trend, with code perfume 6
appearing in nearly five times more submissions than code perfume
8. Consequently, some students have internalised aspects represented
by code perfumes and use them frequently, while those with fewer
occurrences may still be in the process of adopting these
constructs. It is conceivable that positive reinforcement by reporting
code perfumes could support this learning process.


To better understand the limits of the code perfumes defined in this
paper, we contrast this data with the ProgPedia dataset of introductory
assignments, where we only found a total of 14 code perfume instances
(IDs 10, 18, and 19) on 4362 student submissions.  This can be
attributed to the very simplistic nature of those assignments, they
only consist of one file and mostly can be done in 30 to 60 LoC.
While this shows that some students are aware of defensive programming
this early in their programming education, more code perfumes
targetting earlier, simpler stages of coding are clearly needed to
fittingly evaluate students in a CS1/Algorithmics class, while our
code perfumes are more fitted for students in a CS2/Software
Engineering class, as shown above.

\summary{RQ1}{The \toolname identified 8598 instances of code perfumes, highlighting a significant emphasis on defensive programming practices among students, while other conventions are not as commonly embraced.
	Our code perfumes target intermediate learners covering
	design principles; more code perfumes are needed to cover beginners'
	code.}

\subsection{RQ2: Are Code Perfumes Related to Correctness?}


\begin{figure*}
	\centering
	\begin{subfigure}[t]{0.24\linewidth}
		\centering
		\includegraphics[width=\textwidth]{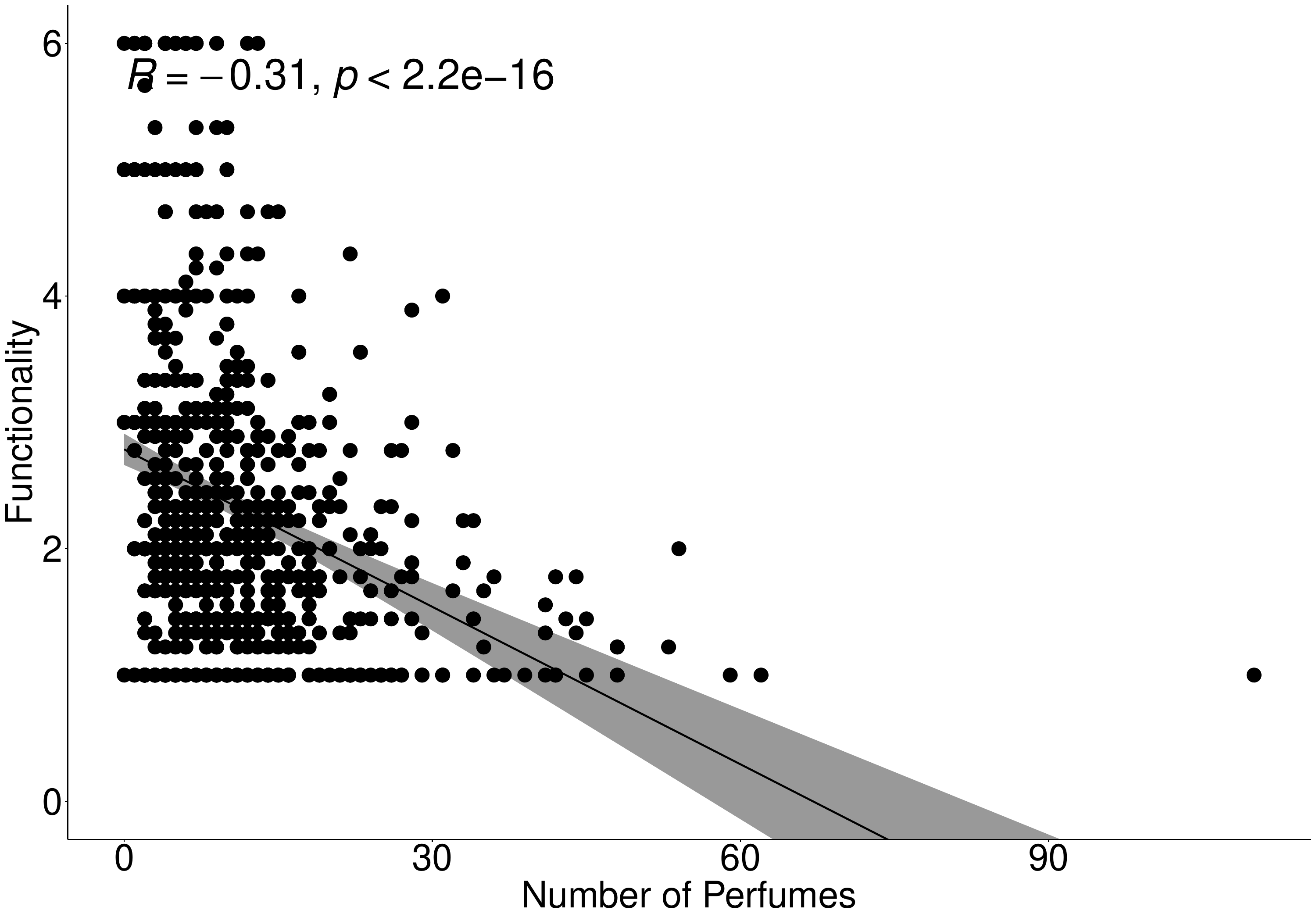}
		\vspace{-1.7em}
		\caption{Number of code perfumes vs. functionality}
		\label{fig:perfumesfunc}
	\end{subfigure}
	\hfill
	\begin{subfigure}[t]{0.24\linewidth}
		\centering
		\includegraphics[width=\textwidth]{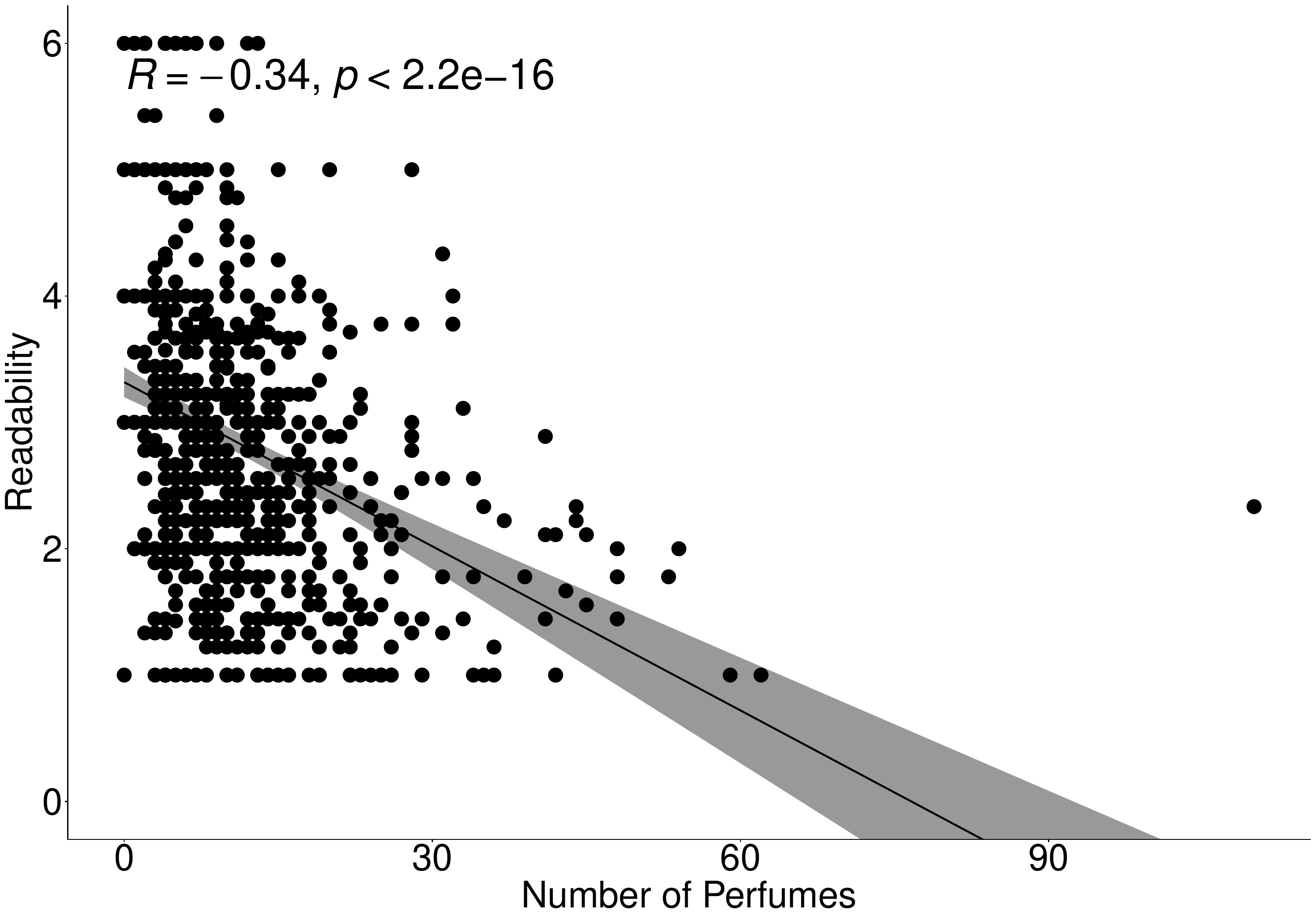}
		\vspace{-1.7em}
		\caption{Number of code perfumes vs. readability}
		\label{fig:perfumesread}
	\end{subfigure}
	\hfill
	\begin{subfigure}[t]{0.24\linewidth}
		\centering
		\includegraphics[width=\textwidth]{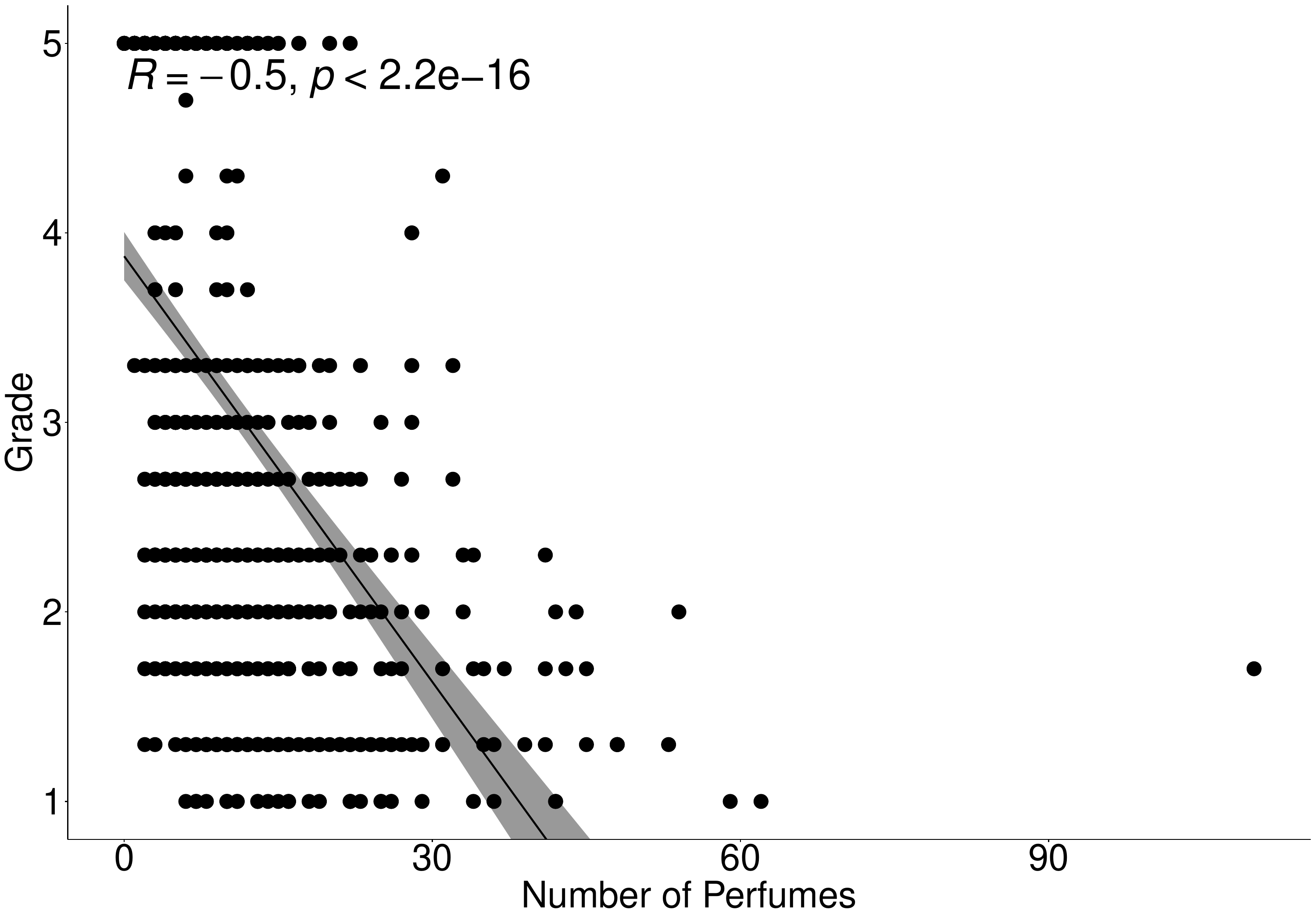}
		\vspace{-1.7em}
		\caption{Number of code perfumes vs. grades}
		\label{fig:perfumesgrade}
	\end{subfigure}
	\hfill
	\begin{subfigure}[t]{0.24\linewidth}
		\centering
		\includegraphics[width=\textwidth]{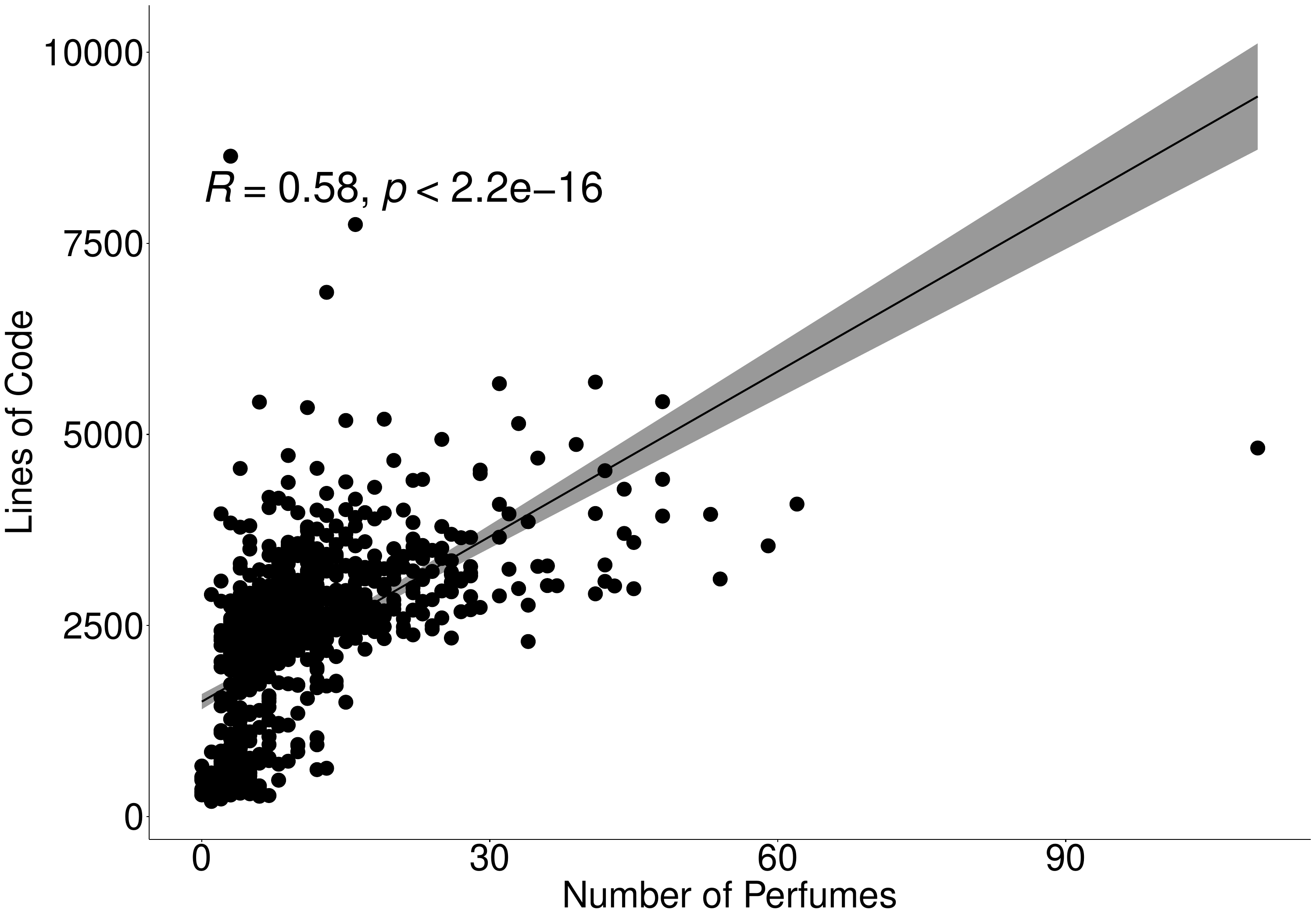}
		\vspace{-1.7em}
		\caption{Number of code perfumes vs. lines of code}
		\label{fig:perfumesloc}
	\end{subfigure}
	\hfill
	\begin{subfigure}[t]{0.24\linewidth}
		\centering
		\includegraphics[width=\textwidth]{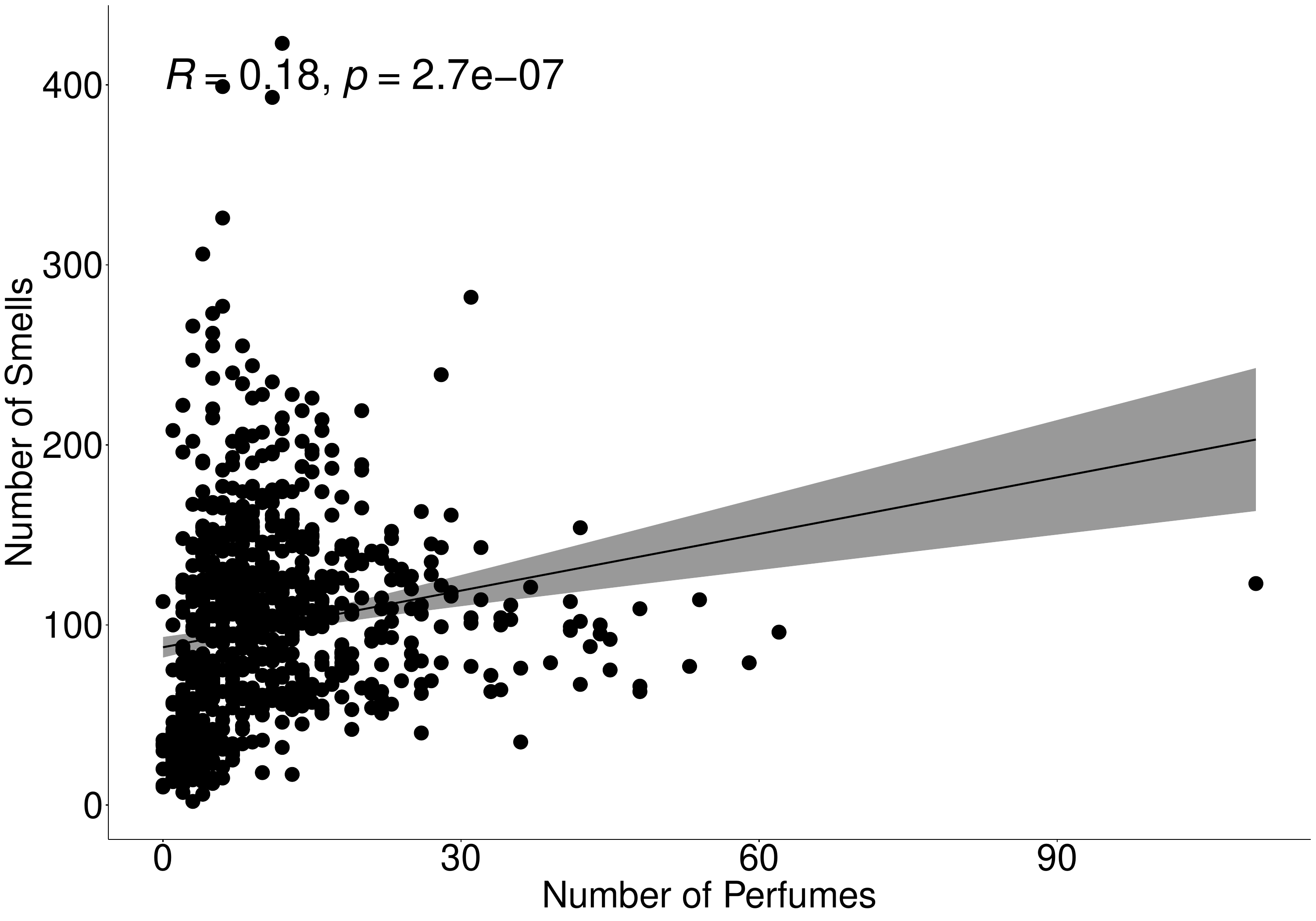}
		\vspace{-1.7em}
		\caption{Number of code perfumes vs. code smells}
		\label{fig:perfumessmells}
	\end{subfigure}
	\hfill
	\begin{subfigure}[t]{0.24\linewidth}
		\centering
		\includegraphics[width=\textwidth]{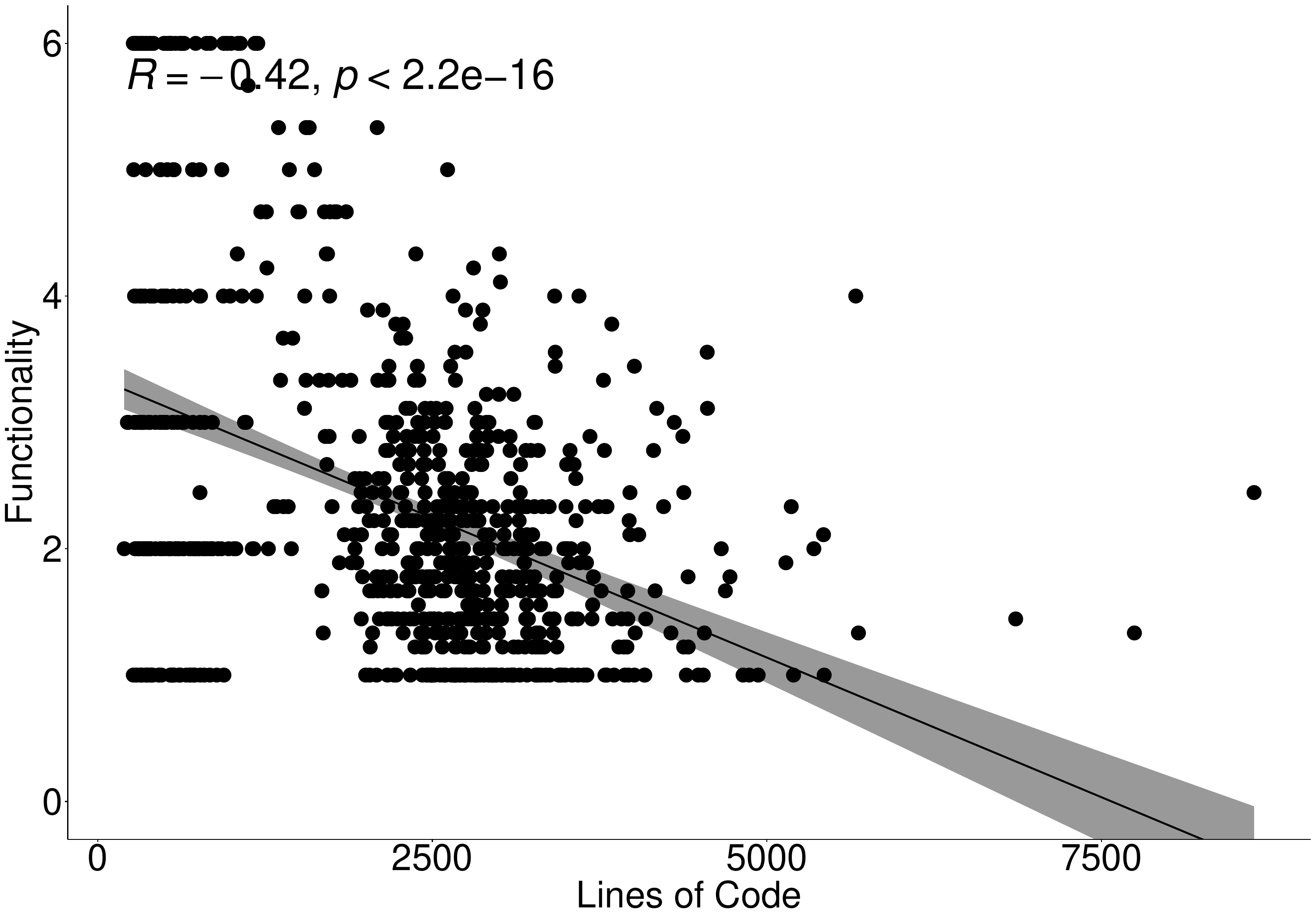}
		\vspace{-1.7em}
		\caption{Lines of code vs. functionality}
		\label{fig:locfunc}
	\end{subfigure}
	\hfill
	\begin{subfigure}[t]{0.24\linewidth}
		\centering
		\includegraphics[width=\textwidth]{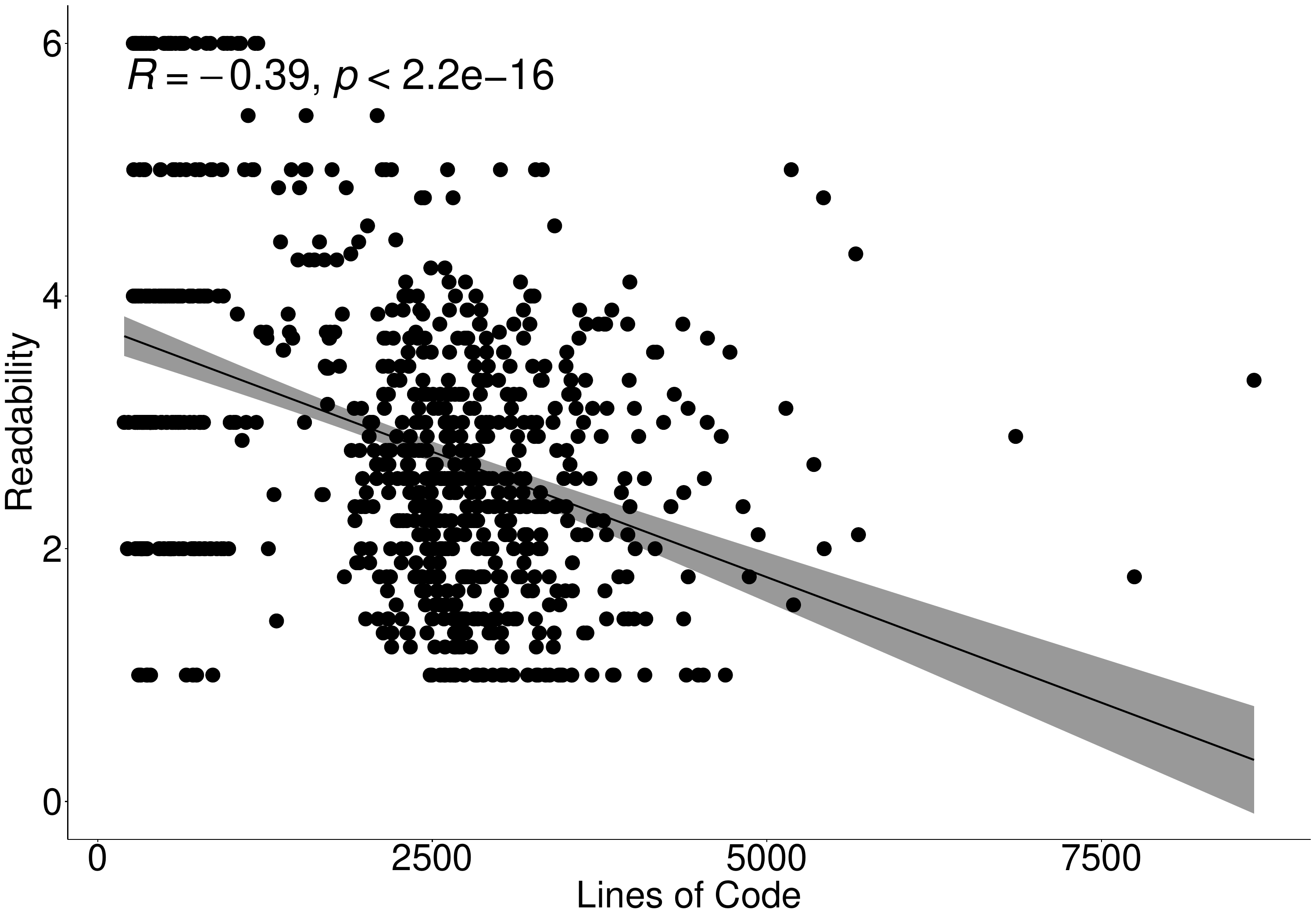}
		\vspace{-1.7em}
		\caption{Lines of code vs. readability}
		\label{fig:locread}
	\end{subfigure}
	\hfill
	\begin{subfigure}[t]{0.24\linewidth}
		\centering
		\includegraphics[width=\textwidth]{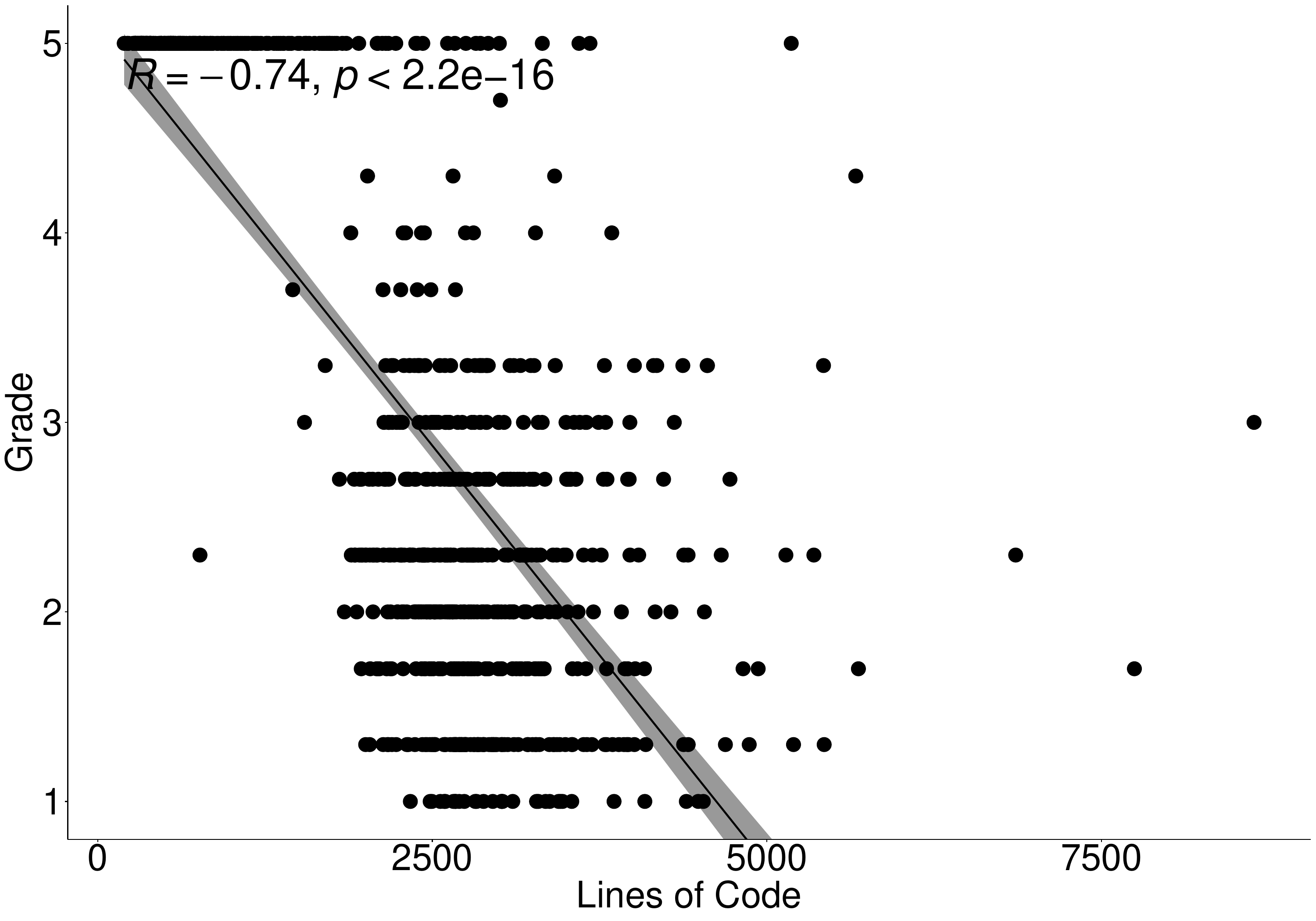}
		\vspace{-1.7em}
		\caption{Lines of code vs. grades}
		\label{fig:locgrade}
	\end{subfigure}
	\hfill
	\begin{subfigure}[t]{0.24\linewidth}
		\centering
		\includegraphics[width=\textwidth]{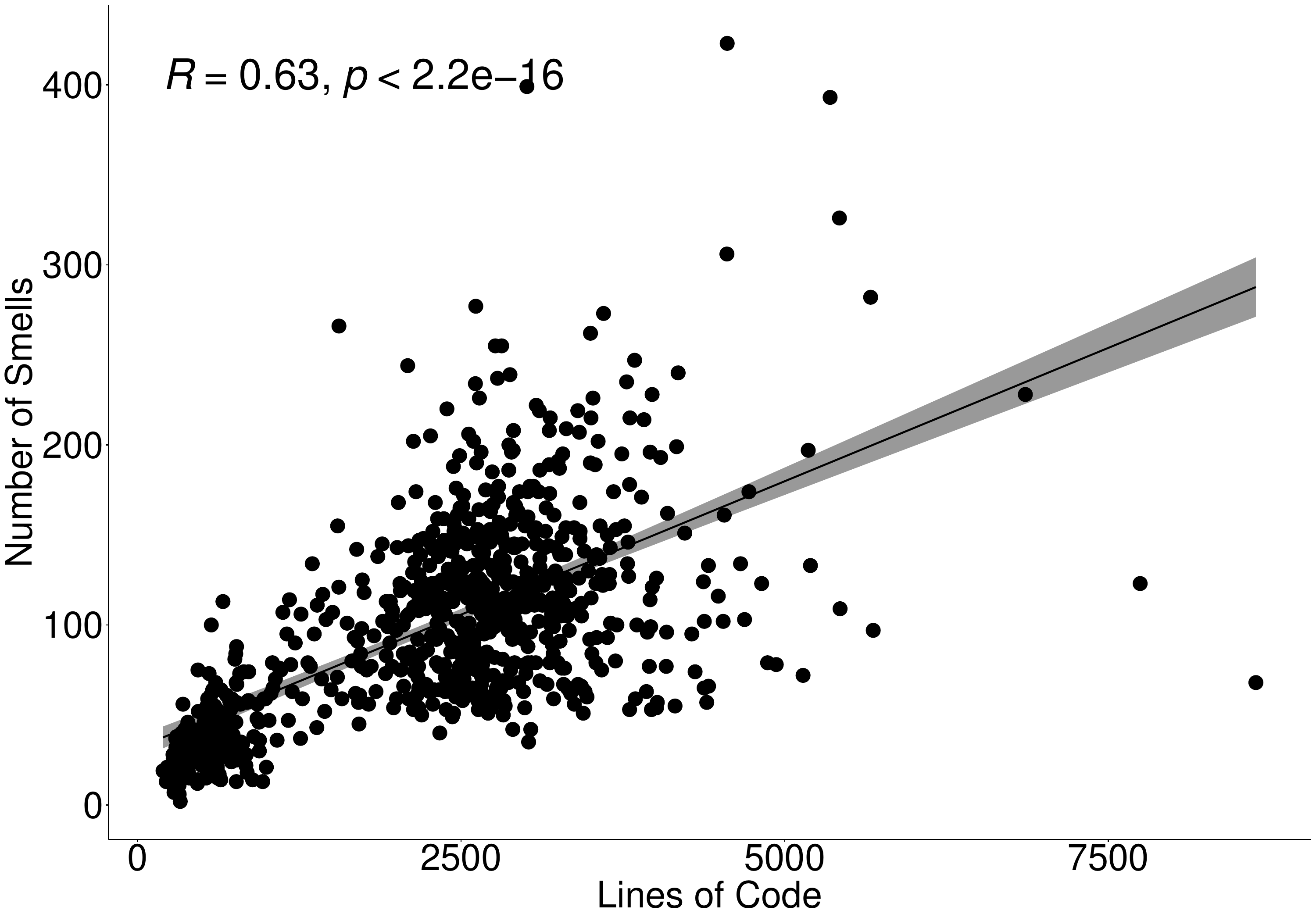}
		\vspace{-1.7em}
		\caption{Lines of code vs. smells}
		\label{fig:locsmells}
	\end{subfigure}
	\hfill
	\begin{subfigure}[t]{0.24\linewidth}
		\centering
		\includegraphics[width=\textwidth]{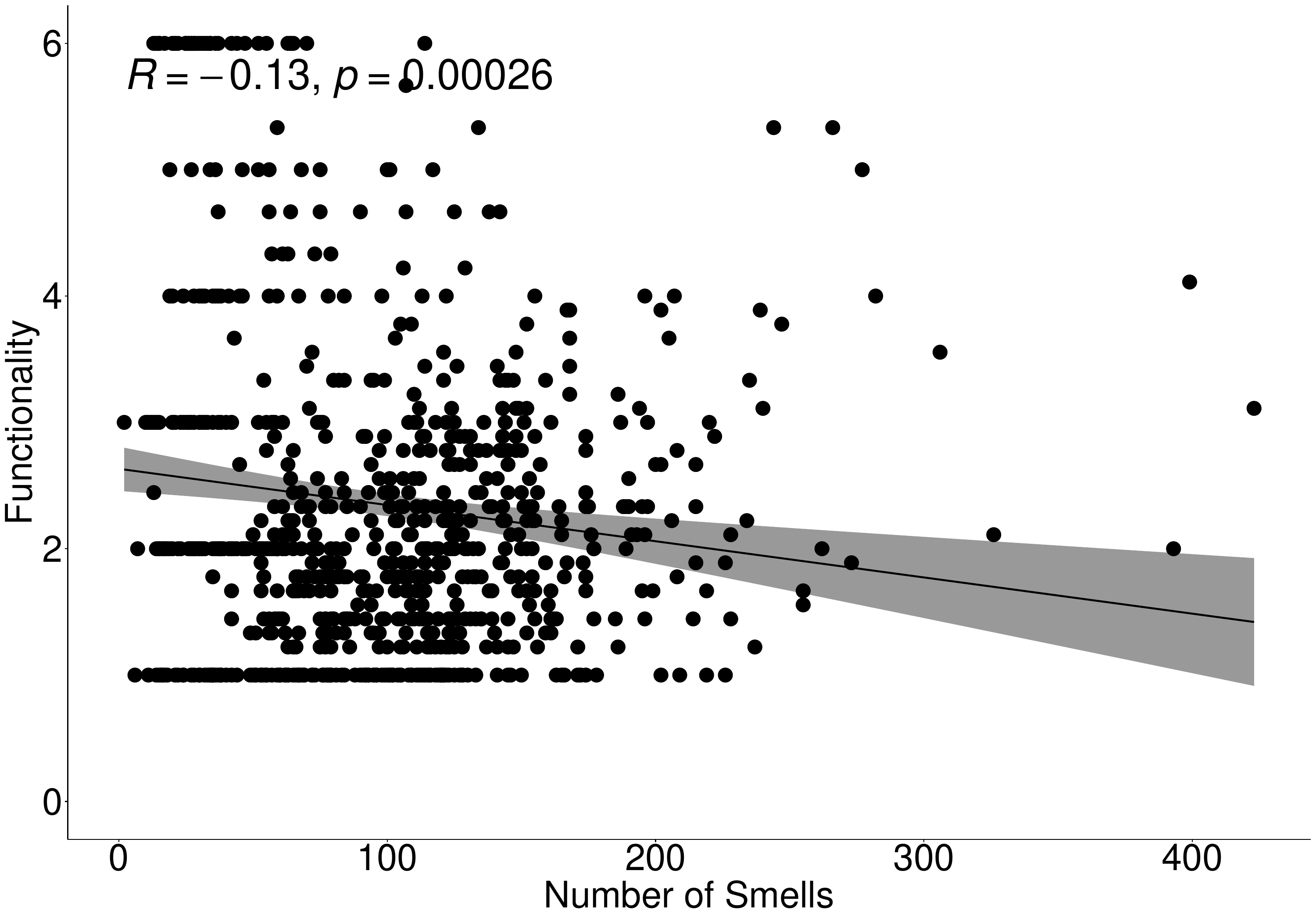}
		\vspace{-1.7em}
		\caption{Number of code smells vs. functionality}
		\label{fig:smellsfunc}
	\end{subfigure}
	\hfill
	\begin{subfigure}[t]{0.24\linewidth}
		\centering
		\includegraphics[width=\textwidth]{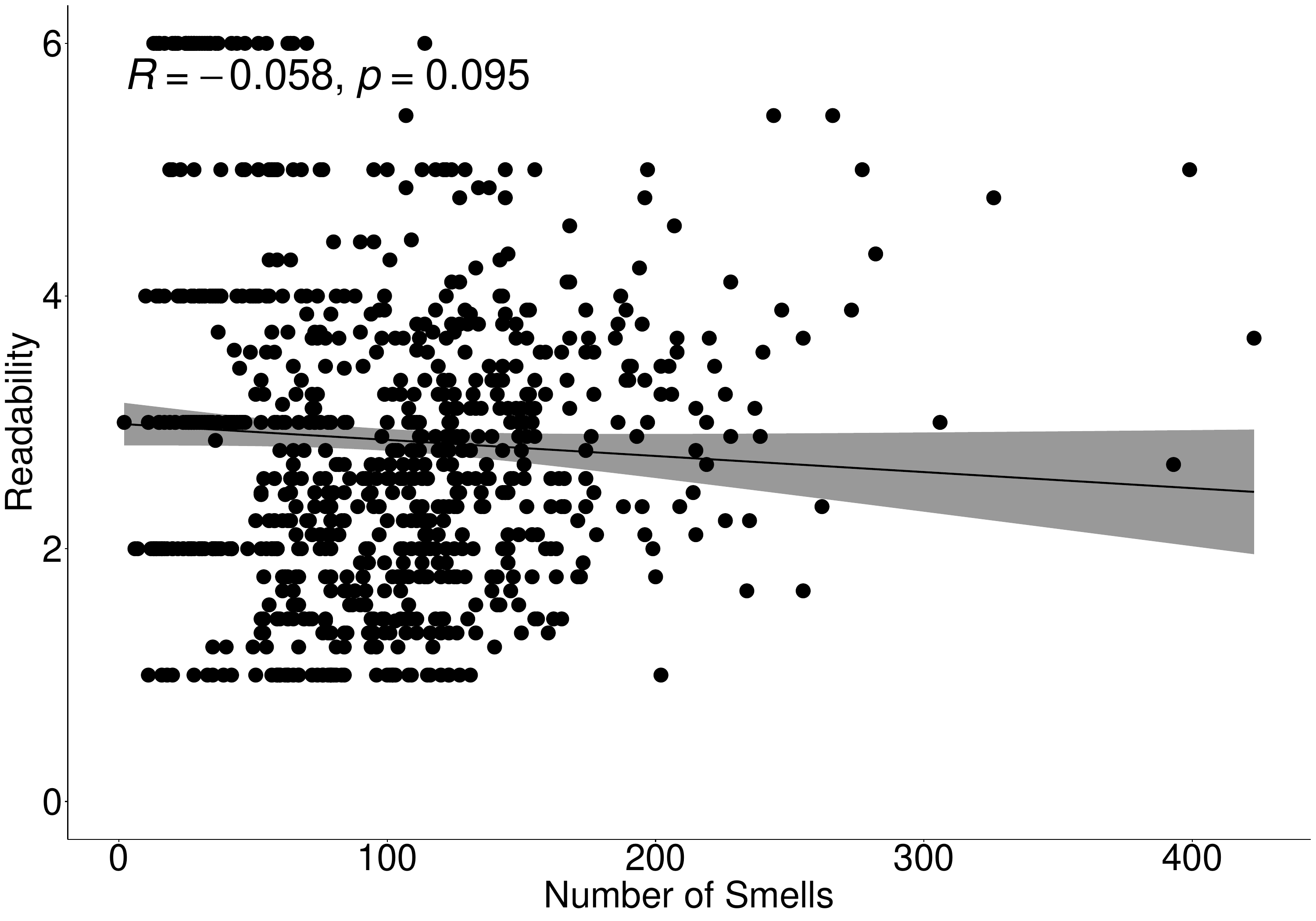}
		\vspace{-1.7em}
		\caption{Number of code smells vs. readability}
		\label{fig:smellsread}
	\end{subfigure}
	\hfill
	\begin{subfigure}[t]{0.24\linewidth}
		\centering
		\includegraphics[width=\textwidth]{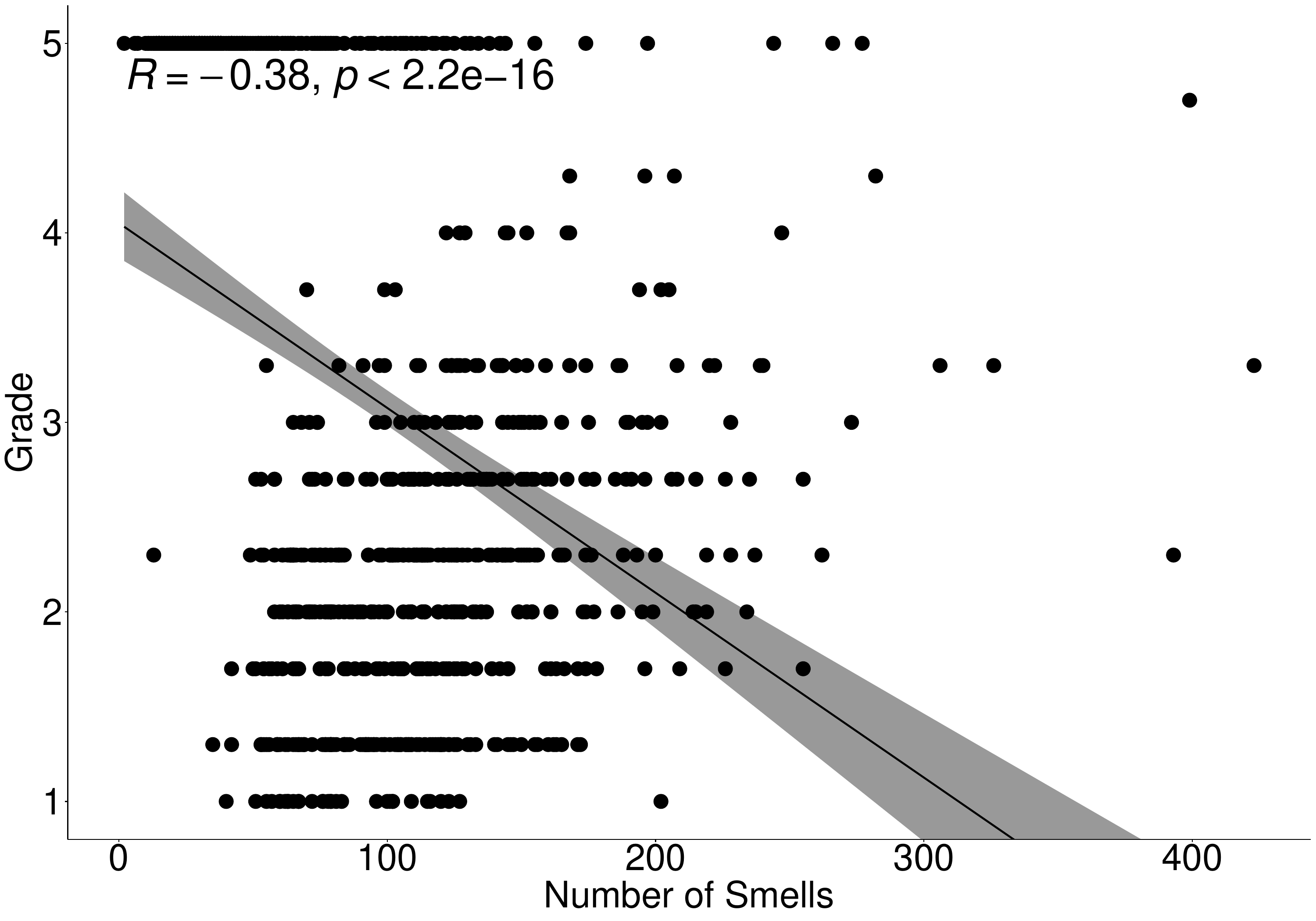}
		\vspace{-1.7em}
		\caption{Correlation between number of code smells and grades}
		\label{fig:smellsgrade}
	\end{subfigure}
	
	\caption{Correlations between number of code perfumes, code smells, line of code, functionality, readability, and grades}
	\label{fig:scorecorrelations}
\end{figure*}

\Cref{fig:scorecorrelations} visualises the correlations between the
metrics considered.
According to Pearson correlation~\cite{cohen2009pearson}, there is a
weak negative correlation between the number of code perfumes and the
functionality (\cref{fig:perfumesfunc}) and readability
(\cref{fig:perfumesread}) assigned to students
assignments. Additionally, there is a moderate negative correlation
between the number of code perfumes and the grades
(\cref{fig:perfumesgrade}) received by the students. Since the best
grade is \enquote{1}, 
the negative correlations suggest that code perfumes can serve as an
indicator for correctness as well as code quality, and thus represent
a valuable tool for educators.


However, we can generally also see that LoC are correlated with
functionality (\cref{fig:locfunc}), readability (\cref{fig:locread}),
and grades (\cref{fig:locgrade}), as more complete solutions tend to
consist of more code. When normalising code perfumes per LoC, we find
no correlations between code perfumes per LoC and functionality
(0.01), readability (-0.04), and grades (0.06). An explanation for
this is that our 20 code perfumes are too few. While code perfumes can
inform tutors about their students and aspects they perform well on,
for code perfumes to serve as a reliable predictor for student
performance, more code perfumes will need to be defined that cover
more aspects of functionality and readability. The immediate
application of code perfumes for educators thus for now primarily lies
in observing which aspects of code are handled well by students, or
identifying students that are lacking understanding of certain
concepts.

The complementary nature of code smells and code perfumes is indicated
by the substantially weaker correlations between code smells and
functionality (\cref{fig:smellsfunc}), readability
(\cref{fig:smellsread}), and grades (\cref{fig:smellsgrade}), even
though more code also leads to more code smells
(\cref{fig:locsmells}). The most common code smells are related to
logging, cognitive complexity, optimising conditionals,
multi-threading and code duplication. The analyses on code smells are
influenced by the large differences between the numbers of code
perfumes and code smells, as SonarLint defines 677
rules~\cite{SonarLintRules}. On the one hand, this reinforces the
conclusion that more code perfumes are needed. On the other hand, it
also suggests that neither mechanism on its own can be used for
reliable grading, but the combination of negative and positive aspects
provides valuable information. Thus, classic linters and the
perfumator could be used in combination to create a more holistic view
of the learners project helping educators support and assess their
learners.


\summary{RQ2}{The presence of code perfumes is associated with
	improved readability, functionality, and higher grades, and
	therefore represent a valuable tool to inform educators about their
	students progress. However, the vast difference in the number of
	existing code smells and code perfumes suggests future work on defining more code perfumes.}

	\section{Conclusions}\label{sec:conclusions}

Program analysis can point out problems in code, but they usually do
not provide encouraging, positive feedback. In this paper, we
therefore introduced the concept of code perfumes for Java, and
presented 20 code perfumes to provide learners with constructive
feedback. Our analysis indicates that many of these code perfumes are
common in practice, and we find a link between the presence of code
perfumes and program correctness, highlighting their potential to aid
teachers' understanding of student learning progress.

While these initial findings are encouraging, our catalogue of 20 code
perfumes is by no means complete, and we anticipate future work to
result in many additional code perfumes.
Investigating how to integrate these in the learning process will be
similarly important. For example, code perfumes could be annotated
directly in Integrated Development Environments (IDEs) or in gamified
environments such as
Gamekins~\cite{DBLP:conf/icse/StraubingerF23}. Educators may need
different ways to interface with code perfumes, for example by
incorporating them as a metric in grading and learning systems such as
ArTEMiS~\cite{DBLP:conf/sigcse/KruscheS18}.
Finally, although our motivation originally was to support learners,
there is no reason why code perfumes could not also be used to
encourage professional programmers.


	The \toolname is available at:
	\begin{center}
		\href{https://github.com/se2p/perfumator}{https://github.com/se2p/perfumator}
	\end{center}
	
	\section*{Acknowledgements}
	\noindent This work is supported by the DFG under grant \mbox{FR 2955/2-1}, 
	``QuestWare: Gamifying the Quest for Software Tests''. Thanks to Jakob 
	Silbereisen for his contributions to the \toolname.
	
	\bibliographystyle{ieeetr}
	\bibliography{bib}
	
\end{document}